\documentstyle[aps,graphicx,preprint]{revtex}
\tightenlines
\draft
\title{Noiseless Collective Motion out of Noisy Chaos}
\author{%
Tatsuo Shibata$^1$,
Tsuyoshi Chawanya$^2$, and
Kunihiko Kaneko$^1$}
\address{%
$^1$%
Department of Pure and Applied Sciences, 
University of Tokyo,
Komaba, Meguro-ku, Tokyo 153-8902, Japan}
\address{%
$^2$%
Department of Physics, Graduate School of Science, Osaka University,
Toyonaka 560-0043, Japan}
\date{November 25, 1998; submitted to Physical Review Letters}

\begin{document}
\maketitle

\begin{abstract}
We consider the effect of microscopic external noise on the collective motion
of a globally coupled map in fully desynchronized states.
Without the external noise a macroscopic variable
shows high-dimensional chaos distinguishable from random motions.
With the increase of external noise intensity,
the collective motion is successively simplified.
The number of effective degrees of freedom in the collective motion
is found to decrease as~$-\log{\sigma^2}$ with the external noise
variance~$\sigma^2$.  It is shown how the microscopic noise can
suppress the number of degrees of freedom at a macroscopic level.
\end{abstract}


Chaotic motions have been observed experimentally
in physical, chemical and biological systems.
Since the evolution of these systems is subjected to external fluctuations,
the observability of deterministic chaos depends on
how the external fluctuations influence on it~\cite{Oono,Crutchefield}.
Motivated by this point,
extensive studies have been carried out
about the enhancement of predictability and unpredictability of
chaotic motions~\cite{Crutchefield,Matsumoto}.

So far such studies are restricted to low-dimensional dynamical systems.
Low-dimensional chaotic motion often arises as a macroscopic motion
out of microscopic chaos with many degrees of freedom.
Let us consider external fluctuations imposed on 
the microscopic level rather than the macroscopic level,
as is probable in natural systems, 
such as fluid turbulence,
or neural systems with a large number of neurons.
Since chaos can amplify a small-scale error,
it would be natural to ask a question
how such a low-dimensional macroscopic chaos is possible 
out of high-dimensional chaotic systems subjected by external fluctuations.

To address the question, we note that
in certain coupled dynamical systems
a macroscopic variable shows seemingly low-dimensional motions,
while microscopic variables keep high dimensional chaos.
Such a phenomenon has been extensively studied
as a collective motion in
coupled map lattice~\cite{CML&CA}, 
globally coupled oscillators~\cite{Oscillator},
and globally coupled map%
~\cite{Kaneko1990b,Pikovsky1994,Perez1993,Shibata1997,Ershov,Chawanya,%
Shibata1998a,Shibata1998b,Nakagawa}.
In the present Letter,
we focus on
the effect of noise on the collective motion of globally coupled map~(GCM).

The present GCM consists of $N$ elements iterated by~a local
dynamics~$f(x)$ with a global coupling among elements and an external
noise. The dynamics is given by
\begin{equation}
x_{n+1}(i) = (1-\epsilon)f(x_n(i))+\frac{\epsilon}{N}\sum_{j=1}^{N}f(x_n(j))+ \xi_n(i),
\label{eq:GCM}
\end{equation}
for the $i$'th element at time step $n$.
Here, we adopt the logistic map $f(x)=1-a x^2$ for
the local dynamics, and Gaussian random process
for~$\xi_n(i)$, with ${\langle\xi_n(i)\rangle=0}$ and
$\langle\xi_n(i)\xi_m(j)\rangle=\sigma^2\delta_{nm}\delta_{ij}$. 
The variance of Gaussian distribution is denoted
by~$\sigma^2$~\cite{CutOff}.

In the noiseless case~($\sigma=0$),
if the coupling strength~$\epsilon$ is small enough, 
the motion of each element seems to be independent from the others.
Even in such cases,
the motion in macroscopic variables counterintuitively does not vanish
in the thermodynamic limit~($N\rightarrow\infty$).
This has been studied as ``collective motion'' in GCM%
~\cite{Kaneko1990b,Pikovsky1994,Perez1993,Shibata1997,Ershov,%
Chawanya,Shibata1998a,Shibata1998b,Nakagawa},
which implies some sort of coherence between elements.

Fig.1 gives an example of the collective motion in GCM~(\ref{eq:GCM})
without the external noise~(${\sigma=0}$).  
We adopt the mean field,
\begin{equation}
h_n={1\over N}\sum_{i=1}^Nf(x_n(i)),
\end{equation}
as a macroscopic observable.
While the microscopic motion shows high dimensional chaos
in the sense that the Lyapunov dimension
is proportional to the number of elements~$N$,
the macroscopic motion shows a quasiperiodic-like structure as is shown in Fig.1.
In almost all the parameter values,
the mean field motion shows some coherent structure
ranging from quasiperiodic-like to higher dimensional one
distinguishable from random motions~\cite{Shibata1998a}.
However,
even if the macroscopic motion looks like quasiperiodic,
scattered points around the torus-like structure depicted in Fig.1
does not vanish even in the thermodynamic limit~\cite{Shibata1998b,Nakagawa},
suggesting high dimensionality of the collective motion.
So far, the mean field motion is considered to be an infinite
dimensional motion even when the torus like structure is 
observed~\cite{Ershov,Chawanya,Shibata1998a,Shibata1998b}.

The addition of noise may, however, destroy such coherence among elements.
One of the authors reported that
the microscopic external noise leads
the mean square deviation~(MSD) of the mean field distribution
decreases with~$N$~\cite{Kaneko1990b}.
(The MSD decreases in proportion to $1/N^{\beta}$ with $\beta\leq1$, 
when $\sigma$ is larger than a certain constant.)
On the other hand, it is also reported that, the external noise sharpens 
the peak in the power spectrum of the collective motion~\cite{Perez1993}.
In this Letter, we clarify the effect of noise on the collective motion in GCM.

Consider a one-body distribution function~$\rho_n(x)$ of the elements
to study the behavior of the collective motion
in the thermodynamic limit~$N\rightarrow\infty$.
Since the mean field value
\begin{equation}
h_n=\int f(x)\rho_n(x)dx,
\label{eq:meanfiled}
\end{equation}
is applied commonly for each element,
and since the additive noise can be represented as
a deterministic diffusion process of the distribution function
in the thermodynamic limit,
the evolution of $\rho_n(x)$ obeys Perron-Frobenius
equation written as
\begin{equation}
\rho_{n+1}(x)=\int dy{1\over\sqrt{2\pi}\sigma}
e^{-{\left(F_{n}(y)-x\right)^2\over2\sigma^2}}
\rho_{n}(y),
\label{eq:PF}
\end{equation}
with
$F_{n}(x)=(1-\epsilon)f(x)+\epsilon h_{n}$.

Fig.2 gives an example of return map of 
the mean field value obtained numerically in GCM with the external noise.
The parameters~$a$ and~$\epsilon$ are the same as in~Fig.1.
Numerical calculation was carried out through integration of
Eq.(\ref{eq:PF})
using a sufficiently large dimensional vector to approximate $\rho_n(x)$.
As is shown in Fig.2(a),
the motion has a clearer structure
than the motion without the noise.
By increasing~$\sigma$,
motions on a torus, locking states and lower dimensional chaos
are observed~(Fig.2(b)).
With the further increase of~$\sigma$,
the collective motion collapses to a fixed point.
Hence, with the increase of the noise a sort of bifurcation to lower
dimensional motions is observed.

To clarify this point,
it seems natural 
to measure
the number of effective degrees of freedom in the collective motion.
We calculate the Lyapunov dimension of the dynamics of $\rho_n(x)$.
The Lyapunov exponents are given by the growth rates of
tangent vectors around the orbit of Eq.(\ref{eq:PF}).
For numerical calculation,
$\rho_n(x)$ is approximated
by a sufficiently large dimensional vector,
and its linear stability around the orbit is studied.

In Fig.3, the Lyapunov dimension denoted by~$D_C$
is plotted as a function of the noise variance~$\sigma^2$.
For sufficiently large~$\sigma$, only the stationary state is
observed and $D_C$ is zero accordingly.
With the decrease of $\sigma$
we have found the low dimensional collective motion~($D_C\sim O(1)$)
ranging from the motion on a torus to low dimensional chaos.
With the further decrease of $\sigma$,
the dimension grows as
\begin{equation}
D_C\propto-\log{\sigma^2}.
\label{eq:dim}
\end{equation}
This implies that the number of effective degrees of freedom
goes to infinity in the zero noise limit,
as is expected from the analysis of the collective motion in GCM
without the external noise.


In the large $\sigma$ regime a variety of bifurcations appears,
which may strongly depend on the parameters.
However, the above result suggests
that the scaling relation (\ref{eq:dim}) will be a characteristic common to
the high-dimensional collective motions
in the small $\sigma$ regime.

Although the evolution rule is originally given for the microscopic
variables, our main interest is on the behavior of macroscopic
variables which would be the only possible observable in typical
cases. Thus, it is highly desirable to obtain a closed description of
the behavior of the macroscopic variables,
which could be written as
\begin{equation}
h_n=h(h_{n-1}, h_{n-2}, \cdots),
\label{eq:funcH}
\end{equation}
for an idealized example.
In most cases, however, it is quite difficult and may well be impossible to
obtain such a description.
Thus, we examine the linear response of the system against
infinitesimal perturbation on the macroscopic variables, and obtain
the variational equation describing the evolution of the small deviation
of the macroscopic variables in a neighborhood of a trajectory.

In the present case, since the elements interact only through the mean
field value, it is quite natural to expect that the behavior of the
mean field value can be consistently described by itself.  
We expect
that the effective number of the dimension of the collective motion
gives substantial agreement with
the Lyapunov dimension of the macroscopic dynamics
estimated in the above mentioned way.
We concentrate on the small~$\sigma$ regime and
give qualitative explanation
for the scaling relation~(\ref{eq:dim}).

If we consider small deviations~$\eta_n$ of~$h_n$,
then $\eta_{n}$ is regarded as a function of $\{\eta_{n-1}, \eta_{n-2}, \cdots\}$.
The evolution of $\eta_n$ from the unperturbed orbit $h_n$ is given by,
\begin{equation}
\eta_n=\sum_{\tau=1}^{\infty}L_\tau\eta_{n-\tau}+O(\eta^2),
\label{EQ:eta}
\end{equation}
where $L_\tau$ is a coefficient
to give the linear response of the mean field value
at $n$ step to the displacement at ${n-\tau}$ step.
The number of the Lyapunov dimension of the mean field dynamics
is estimated from the eigenvalues of this linear regression~\cite{lin}.

First we estimate $L_\tau$ from the dynamics of the distribution function
given by Eq.(\ref{eq:PF}).
In the small noise limit~($\sigma~\rightarrow~0$),
from Eqs.(\ref{eq:meanfiled})~and~(\ref{eq:PF}),
$L_\tau$ is given by
\begin{equation}
L_\tau  =  {\epsilon\over1-\epsilon}
\int dx{d F^{(\tau)}_{n}(x)\over dx}
\rho_{n-\tau+1}(x),
\label{eq:hint}
\end{equation}
where
$F^{(\tau)}_{n}(x)\equiv F_{n}\circ F_{n-1}%
\circ\cdots\circ F_{n-\tau}(x)$.

For $\tau\gg1$,
${dF^{(\tau)}_n(x)/dx}$ in Eq.(\ref{eq:hint})
changes its sign quite frequently in $x$.
Now let us consider the partition of $x$ at the points such that
$F^{(\tau)}_n(x)=0$.
Denoting the typical value of
${\left|dF^{(\tau)}(x)/dx\right|}$ by $d(\tau)$,
the interval of partitions is estimated at $1/d(\tau)$,
which decreases rapidly with $\tau$.
Since the integration in a partition becomes zero if $\rho(x)$ stays
constant in that partition,
the partitions where $\rho_n(x)$ changes drastically in $x$
contribute to the estimation of $L_\tau$
much more than
the partitions where $\rho_n(x)$ does not changes so much.

In the case of small noise limit~($\sigma\rightarrow0$),
the most drastic change of $\rho_n(x)$
comes from the inverse square-root singularities,
which is the characteristic structure of distribution function for
the logistic map.
Hence, the integration in the partitions
containing the characteristic structure in $\rho_n(x)$
is estimated as $O(\sqrt{d(\tau)})$~\cite{integration}.
$d(\tau)$ is roughly estimated at $e^{\lambda_m\tau}$ for $\tau\gg1$,
where $\lambda_m$ is the Lyapunov exponent of the local mapping~\cite{fluctuation}.
Consequently,
the response $L_\tau$
to the perturbation grows exponentially
with the rate~${1\over2}\lambda_m$.

Even in the presence of finite amplitude of the noise,
the above order estimation for $L_\tau$ is still valid
for $\tau$ smaller than $\tau_c\equiv -\log{\sigma}/\lambda_m$,
where the typical width of the partitions becomes
comparable with the typical amplitude of the noise, 
i.e. $1/d(\tau)\sim e^{-\lambda_m\tau}=\sigma$.

For larger $\tau>\tau_c$, however,
the effect of noise in smoothening the distribution~$\rho(x)$
appears so that $L_\tau$ will start to decay with $\tau$.

Partially integrating (\ref{eq:hint}), 
we obtain
\begin{eqnarray}
L_\tau  &=&  -{\epsilon\over1-\epsilon}
\int dxF^{(\tau)}_{n}(x){d\rho_{n-\tau+1}(x)\over dx}\cr
 &=&  -{\epsilon\over1-\epsilon}
\int dx\tilde F^{(\tau)}_{n}(x){d\rho_{n-\tau+1}(x)\over dx},
\label{eq:hint2}
\end{eqnarray}
with $\tilde F^{(\tau)}_{n}(x) = F^{(\tau)}_{n}(x) - \overline{F}^{(\tau)}_{n}$,
where $\overline{F}^{(\tau)}_{n}$ is the average value of $F^{(\tau)}_{n}(x)$
over the support of $\rho_n(x)$.
With the increase of $\tau$,
$\tilde F^{(\tau)}_{n}(x)$ becomes
a rapidly oscillating function about zero-mean in $x$,
and the integration of $\tilde F^{(\tau)}_{n}(x)$
over any finite range within the support of $\rho_n(x)$
will approach zero~\cite{mixing}.
Since $|{d\rho_{n-\tau+1}(x)\over dx}|$ is uniformly bounded
due to the existence of the noise, 
$L_\tau$ converges to zero.
Hence, as far as the linear stability is concerned,
the mean field value is not sensitive
to the mean field values before sufficiently long time step~(before
$O(\tau_c)$ steps).
Thus we can consider a dynamics of $h_n$
as a function of the mean field values of the past $O(\tau_c)$ steps
as Eq.(\ref{eq:funcH})
at least in the neighborhood of the orbit.

In summary,
the amplitude of $L_\tau$ grows with $\tau$ as
$L_\tau\sim e^{{1\over2}\lambda_m\tau}$ for $\tau<O(\tau_c)$
whereas it starts to decay for $\tau>O(\tau_c)$.
Thus for sufficiently large $\tau_c$,
i.e., for sufficiently small $\sigma$,
the number of positive eigenvalues around zero for
Eq.(\ref{EQ:eta}) is estimated at $O(\tau_c)$.
Since we have to consider the contribution
only from the latest $O(\tau_c)$ steps,
the dimension of this dynamical system
can be at most $O(\tau_c)$.
Hence, the dimension of the mean field dynamics
is within the order of $\tau_c$.
Accordingly
the number of effective degrees of freedom of the mean field dynamics
grows as $-\log{\sigma}$ with the decrease of $\sigma$,
and can grow arbitrary large as $\sigma$ approaches zero.

In the present Letter,
we have shown that the noise in a microscopic level
reduces the complexity of the collective motion,
which is characterized by the number of degrees of freedom.
It is shown numerically that
the dimension of the dynamics of Perron-Frobenius equation~Eq.(\ref{eq:PF})
satisfies (\ref{eq:dim}).
On the other hand,
analysis on the mean field dynamics also supports (\ref{eq:dim}). 
Hence,
the number of effective degrees of freedom of the collective motion
in the present GCM is
concluded to satisfy the scaling relation (\ref{eq:dim}).

Such a relation is expected to hold
when the collective motion keeps high dimensional motion
with microscopic chaos.
High-dimensional collective motions are supported by the
exponential growth of the linear response coefficient.
When the distribution has peak concentrations
within a small width $\delta x$,
such an exponential growth is expected
until $\tau\sim\log{\delta x}$ 
and thus a high dimensional collective motion appears~\cite{expgr}.
Even when the local dynamics is given as a higher-dimensional
non-hyperbolic map or a chaotic flow,
the present argument on the collective motion is
expected to hold.
Hence, the logarithmic dependence of the dimension upon the noise
intensity given by~(\ref{eq:dim})
will be observed in a broad range of systems.

The induced regularity by the addition of noise
was also reported as Noise-Induced Order~(NIO)
in a one-dimensional map~\cite{Matsumoto}.
The mechanism to induce regularity in our system has
a similarity with NIO case, in the point
that the external noise
destroys a dynamical structure
which causes the irregular behavior.
In NIO,
the noise reduces the concentration of measure
in the instability region where the intermittent behavior is generated.
On the other hand, in the present case,
the noise smoothes the singularity of the distribution function,
that is the source of high-dimensional instabilities
of the collective motion.

In the case of NIO, however,
the motion is not completely regular, 
since the noise is imposed upon the observed variable itself.
On the other hand,
in the present case,
the noise is imposed on the microscopic variables,
whereas observed is the macroscopic collective variable.
Thus, the noise-induced motion is {\it deterministic}
and low dimensional in the thermodynamic limit.

We should also mention that
our present result may be applicable to experimental systems,
such as fluid turbulence
and neural systems consisting of a huge number of neurons
with nonlinear dynamics.
By controlling the thermal noise or some external fluctuations,
we expect to observe the gradual reduction of the dimension and noise-induced
low-dimensional motions,
and hence the existence of the collective chaos will be clarified.

The authors are grateful to
N.~Nakagawa, S.~Morita, T.~Hondou, S.~Sasa, K.~Sekimoto, A.~Vulpiani,
H.~Chat\'e, Y.~Kuramoto and T.~Tsuzuki 
for stimulating  discussions.
TS also thanks 
Yukawa Institute for Theoretical Physics at Kyoto University
for his residence,
during which part of this manuscript was written.
This work is partially supported by
Grant-in-Aids for Scientific Research
from the Ministry of Education,
Science, and Culture of Japan.


\begin{figure}
\caption{A return map of the collective motion 
in GCM~(\ref{eq:GCM}) without noise. 
$a=1.86, \epsilon=0.1, \sigma=0.0, N=10^7.$}
\end{figure}

\begin{figure}
\caption{A return maps of the collective motion in
GCM~(\ref{eq:GCM}) with noise. 
$a=1.86, \epsilon=0.1$. The noise variances are 
(a)~$\sigma^2=1.5\times10^{-6}$, and
(b)~$\sigma^2=3.0\times10^{-6}$.
Numerical calculation was carried out with integration of
Eq.(\ref{eq:PF}) using a sufficiently large dimensional vector to
approximate the distribution function.}
\end{figure}

\begin{figure}
\caption{The dimension of the collective motion is plotted as a
function of noise intensity, obtained by the Lyapunov 
dimension. The Lyapunov exponents are calculated as growth rates of
the tangent vector around the orbit, obtained by Eq.(\ref{eq:PF}).
For numerical calculation, 
$\rho_n(x)$ is approximated by a sufficiently large dimensional vector,
and the tangent vectors are orthonormalized at each time step.}
\end{figure}

\noindent\includegraphics{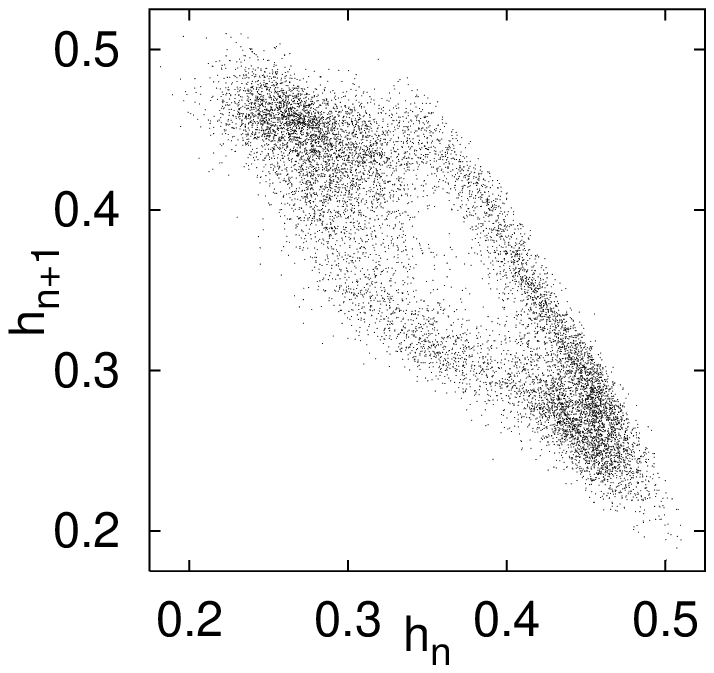}

\noindent\includegraphics{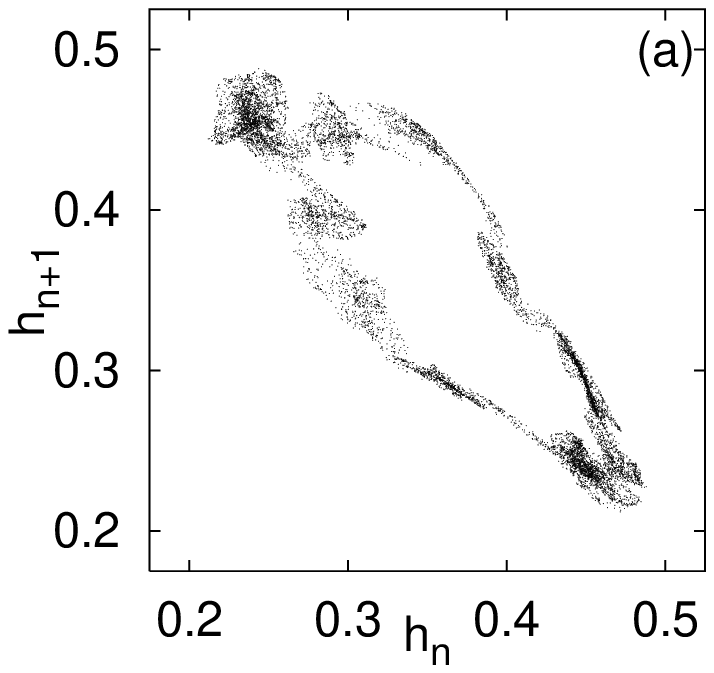}

\noindent\includegraphics{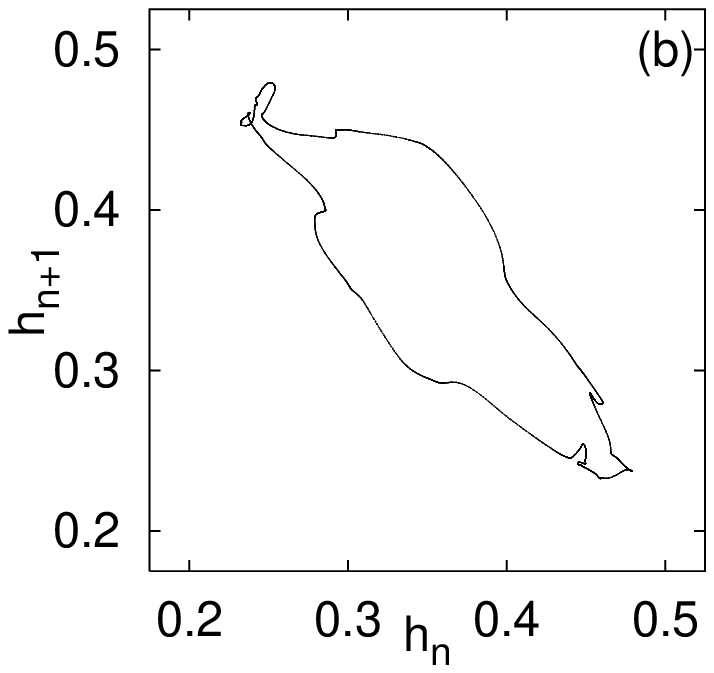}

\noindent\includegraphics{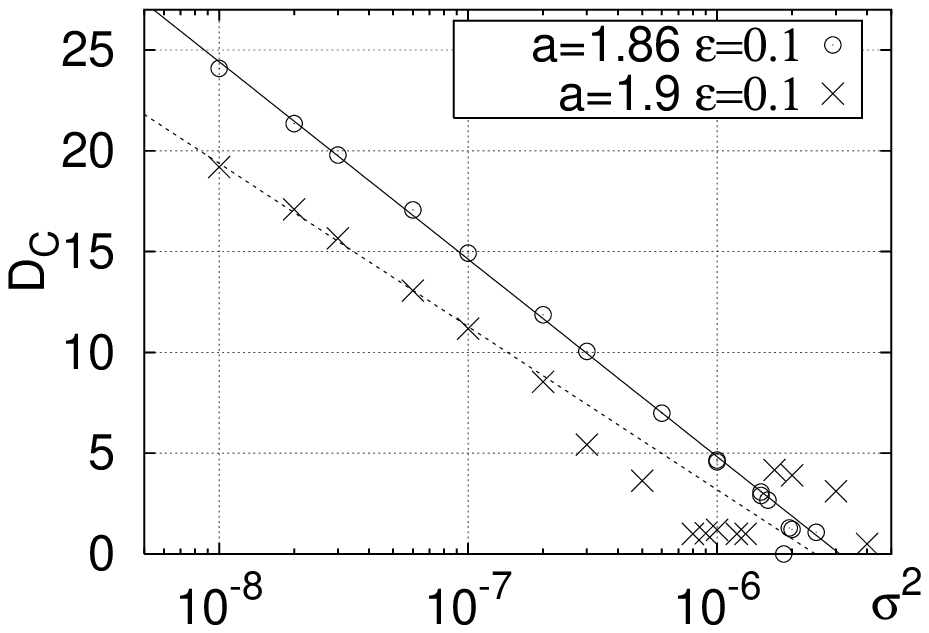}


\begin{thebibliography}{99}
\bibitem{CutOff}
In order to bound the system within a finite region,
``cut off'' of $\xi$ is introduced, so that $f(f(1+\xi)+\xi)>f(1+\xi)$.
In this Letter,
we study small $\sigma$ regime,
where the cut off effect is practically unnecessary for numerical calculation.
\bibitem{lin}
This does not hold
if the effects of perturbations at different time steps
could cancel out completely,
as in the case that
$\rho_n(x)$ contracts into a $\delta$-function.
In the present case, however,
as long as $\rho_n(x)$ has a continuous support,
as is expected from the microscopic high-dimensional chaos,
the effect of $\eta_n$ does not disappear completely by the
perturbation at the other time step.
\bibitem{mixing}
Here we assume the map $F^{(\tau)}_n(x)$ has mixing property,
which seems natural when
$\rho_n(x)$ has a continuous support.
\bibitem{fluctuation}
Here we neglect the fluctuation of the expansion rate.
\bibitem{expgr}
Such a structure is generic in the GCM with logistic type
map~${f(x)=1-a|x|^\alpha}$~($\alpha\geq1$),
where the high dimensional collective motions
is observed~\cite{Ershov}.
\end{thebibliography}
\end{document}